\documentclass[fleqn,usenatbib]{mnras}

\usepackage{newtxtext,newtxmath}

\usepackage[T1]{fontenc}
\usepackage{ae,aecompl}
\usepackage{hyperref}	
\hypersetup{colorlinks=true,linkcolor=blue,citecolor=blue,filecolor=blue,urlcolor=blue}

\usepackage{natbib}
\usepackage{chngpage}
\usepackage{soul}
\usepackage{ulem}
\usepackage{color}
\usepackage{wasysym}
\usepackage{amssymb}
\usepackage{graphicx}
\usepackage{float}
\usepackage[utf8]{inputenc}

\usepackage{threeparttable}

\def\la{\raise.5ex\hbox{$<$}\kern-.8em\lower 1mm\hbox{$\sim$}}
\def\ga{\raise.5ex\hbox{$>$}\kern-.8em\lower 1mm\hbox{$\sim$}}
\def\be{\begin{equation}}
\def\ee{\end{equation}}
\def\ba{\begin{eqnarray}}
\def\ea{\end{eqnarray}}

\def\Omegastar{\Omega_\ast}
\def\OmegaK{\Omega_{\mathrm{K}}}

\def\Mdotin{\dot{M}_{\mathrm{in}}}

\def\Mdot{\dot{M}}

\def\Pdot{\dot{P}}

\def\Pddot{\ddot{P}}
\def\Msun{M_{\astrosun}}

\def\rin{r_{\mathrm{in}}}
\def\rlc{r_{\mathrm{LC}}}

\def\rco{r_{\mathrm{co}}}

\def\Lx{L_{\mathrm{x}}}

\def\rA{r_{\mathrm{A}}}

\def\dM*{\delta M_*}

\def\P0min{P_{0,{\mathrm{min}}}}

\def\Alfven{Alfv$\acute{e}$n}

\def\rbb1{R_{\mathrm{BB1}}}
\def\rbb2{R_{\mathrm{BB2}}}

\def\bt{\begin{tabular}{@{}c@{}}}
\def\et{\end{tabular}}

\def\rM{r_{\mathrm{M}}}

\def\4U{4U 1626--67}

\begin{document}
\title[]{On the peculiar torque reversals and the X-ray luminosity history of the accretion-powered X-ray pulsar 4U 1626--67}

\author[Benli]{
O. Benli\thanks{E-mail:onurbenli@sabanciuniv.edu}
\\
Department of Physics and Astronomy, Southampton University, Southampton SO17 1BJ, UK \\
Observatoire Astronomique de Strasbourg, 11, rue de l'Universit\'e, 67000 Strasbourg, France
}

\date{Accepted XXX. Received YYY; in original form ZZZ}
\pagerange{\pageref{firstpage}--\pageref{lastpage}} \pubyear{----}

\label{firstpage}
\maketitle

\begin{abstract}

We have investigated the rotational and the X-ray luminosity ($\Lx$) observations of 4U 1626--67 accumulated during the last four decades. It has been recorded that the source underwent a torque reversal twice. We have tried to understand whether this eccentrical sign-switch of the period derivative ($\Pdot$) of the source could be accounted for with the existing torque models. We have found that the observed source properties are better estimated with the distances close to the lower limit of the previously predicted distance range ($5-13$ kpc). Furthermore, assuming an inclined rotator, we have considered the partial accretion/ejection from the co-rotation radius that leads different $\Lx$-$\Pdot$ profiles than the aligned rotator cases. We have concluded that the oblique rotator assumption brings at least equally best fitting to the observed X-ray luminosity and the rotational properties of \4U. More importantly, the estimated change of the mass accretion rate which causes the change in observed $\Lx$ of \4U is much less than that is found in an aligned rotator case. In other words, without the need for a substantial modification of mass accretion rate from the companion star, the range of the observed X-ray luminosity could be explained naturally with an inclined magnetic axis and rotation axis of the neutron star.

\end{abstract}



\section{Introduction}\label{sec:intro}

In accreting X-ray binaries where the period change of the compact star is observed to be steady and persistent, the accretion on to the star is thought to occur along with the disc. Low Mass X-ray Binaries (LMXBs) are good examples of those kinds of systems. On the other hand, in the High Mass X-ray Binaries (HMXBs), the mass accretion is fed mostly by the wind from the companion star. In such a wind-fed accreting system, period derivative varies in short timescales because of the quick alternation of accretion geometry and variation in the amount of matter being accreted. Also, short but steady episodes of spin-change observed from some wind-fed systems, are indicative of the formation of the short-duration transient accretion discs in these systems.

Some accretion powered X-ray pulsars have shown fascinating torque reversals separated by time-scales from weeks to months while for some systems, this time-scale can be as much as a few decades. The X-ray pulsar 4U 1626--67 in a LMXB was first observed by \textit{Uhuru} \citep{giacconi+_72}. The pulsar with the spin period $P \simeq 7.6$~s \citep{rappaport+_77}, was confirmed to reside in an ultra-compact binary system with a very-low-mass companion with the estimated mass in the $\sim 0.03 - 0.09 \Msun$ range. The actual mass of the companion varies depending on the binary inclination angle $11^{\circ} \leq i \leq 36^{\circ}$ \citep{levine+_88, chakrabarty+_97}. An optical counterpart, KZ TrA, was subsequently detected by \citep{mcclintock+_77}. Pulsed optical emission which was attributed to reprocessing of pulsed X-rays from the surface of the companion, revealed the 42 min. orbital period of the binary \citep{middleditch+_81, chakrabarty_98}. The existence of an accretion disc around \4U has been well established via particularly double-peaked emission lines in the X-ray band \citep{schulz+_01}. 

The torque history of 4U 1626--67 was reviewed in detail by \cite{camero-arranz+_10}. For the recent spectral and timing analysis of the source, we refer to \cite{iwakiri+_19}. The persistent X-ray pulsar that has been monitored various times since its discovery, underwent two torque reversals on June 1990 \citep{chakrabarty+_97} and February 2008 \citep{camero-arranz+_10}. The source was steadily spinning-up (SU) with the period derivative $\Pdot \sim 10^{-10}$~s~s$^{-1}$ and the X-ray luminosity $\Lx \sim 10^{37}$~erg~s$^{-1}$ before its transition to the spin-down (SD) phase in 1990. Interestingly, the source was spinning down with a similar rate to the pre-reversal SU rate. Subsequently, the source was observed many times in the steady SD phase for 18 years until the second torque reversal occurred in 2008, switching again to the SU phase. 

The elaborated observations of 4U 1626--67 indicate that the torque acting on the neutron star and the X-ray luminosity are related to each other in most cases. The source was able to monitor in the course of the latter reversal. On the contrary, there was no observation during the first reversal. However, observation of the source in 1991 revealed that the pulsar started to SD which confirmed the sign change of the pulsar's $\Pdot$. \cite{camero-arranz+_10} showed that the X-ray flux of the source in 2010, two years after the torque reversal in 2008, has reached up almost the same level as in 1977. The 2008-reversal occurred at a similar but slightly lower X-ray luminosity level compared to the 1990-reversal.

\cite{chakrabarty+_97} estimated that the mass accretion rate on to the star is supposed to be greater than $4 \times 10^{16}$~g~s$^{-1}$ ($2 \times 10^{-10} \Msun$~yr$^{-1}$) for the observed period derivative, $\Pdot = - 6.5 \times 10^{-12}$~s~s$^{-1}$ ($\dot{\nu} = 8.5 \times 10^{-13}$~Hz~s$^{-1}$) at that time. \cite{chakrabarty_98} favored an X-ray irradiated accretion disc over the radiatively driven wind accretion and showed that irradiated disc calculation provides a better fit to the optical photometry. 

Much longer spin-up timescales of Cen X-3 and Her X-1, compared to those of other accreting systems, were not in line with the initiative models proposed to explain spin behavior of accreting pulsars \citep{rappaport_joss_77, ghosh_lamb_79}. Cen X-3 and Her X-1 also showed short-term spin-down behavior interspersed among their spin-up epochs. The X-ray luminosity of these sources did not vary significantly during transitions indicating that the sources continued to accrete while the torque acting on the sources was observed to change its sign. Therefore, the X-ray luminosity and torque variations of Cen X-3 and Her X-1 have not been matching with the prediction of existing accretion models so far.

We tested here whether the X-ray luminosity and the torque ($\Pdot$) history of 4U 1626--67 recorded in the last $\sim 40$~years could be reproduced by the different regimes in the accretion disc-torques. We aimed to account for the $\Lx$ and the $\Pdot$ of the source by considering the simultaneous effect of the spin-up and the spin-down torques acted on the star by an accretion disc as well as taking into account oblique rotator effects. We employed two different torque models and implemented the inclination of the magnetic axis of the star to its rotation axis into one of these models. In Section \ref{sec:accretion}, we describe the models. The results are presented in Section \ref{sec:results}. We discuss the implication of our results and draw the conclusion in Section \ref{sec:dis_conc}.

\section{Accretion on to magnetised star}\label{sec:accretion}

In a Keplerian accretion disc, the angular velocity of the falling matter, $\OmegaK$, increases as it becomes closer to the star. At the co-rotation radius, $\OmegaK$ is equivalent to the angular velocity of the star, $\OmegaK(\rco) = \Omegastar$ which can be written as

\be
r_{\mathrm{co}}= \left(\frac{G M} {\Omega_{*}^{2} } \right)^{1 / 3} = \left(\frac{G M P^{2}}{4 \pi^{2}}\right)^{1 / 3}.
\ee

The inner radius of the accretion disc, $\rin$, is usually expressed as a fraction of the conventional \Alfven~radius,  

\be
r_{\mathrm{A}}=\left(\frac{\mu^{2}}{\sqrt{2 G M} \dot{M}}\right)^{2 / 7}
\ee

where $M$ and $\mu$ are the mass and magnetic dipole moment of the star, $G$ the gravitational constant and $\Mdot$ the mass inflow rate at that radius. In the propeller phase, the realistic inner disc radius where viscously inflowing disc truncated by the dipole field of the star is expected to be close to $\rA$. In literature, the relation $\rin = (0.5-1)~\rA$ has been widely used to define $\rin$. The actual location of $\rin$ is dependent on the details of disc-filed interaction \citep{ghosh_lamb_79, arons_93}. Some authors assume that the critical value of $\rA$ below which accretion onto star is expected to be switched-on is determined by the imaginary light cylinder radius $\rlc = c / \Omegastar$. When $\rA < \rlc$, the system is assumed to be in the accretion where the disc matter arriving at $\rin$ flows onto the star. However, the observed rotational properties and X-ray luminosity of transitional millisecond pulsars contradict with the condition that $\rA$ has to be closer or larger than $\rlc$ to switch-off accretion onto the star (Archibald et al. 2015, Papitto et al. 2015, Ertan 2017).       

In the accretion phase, the X-ray luminosity of the neutron star is calculated by

\be \label{eq:Lx}
L_{\mathrm{X}}=\frac{G M \dot{M}_{\mathrm{acc}} }{R}
\ee

where $\dot{M}_{\mathrm{acc}}$ is the mass accretion rate onto the star and $R$ is the radius of the star. Accretion on to neutron star through a prograde disc makes the star spinning faster. Because the angular momentum of the matter coupled to the magnetic dipole field is superimposed over that of the star even if the total mass of the accreted matter is negligible in reference to the mass of the star. In other words, the neutron star spins up due to the torque by the material being accreted \citep{pringle+_72},

\be \label{eq:su_torque}
\Gamma_{\mathrm{SU}} =\dot{M}_{\mathrm{acc}} \sqrt{G M \rin}
\ee

In addition to the spin-up material torque, spin-down torque also acts on the star due to the interaction between the dipole field and plasma in the disc \citep{ghosh_lamb_79, wang_95, rappaport+_04, erkut_alpar_04, ertan_erkut_08}. Earlier theoretical models imply that neutron stars with conventional dipole field strengths could be in the fast-rotator phase while at the same time continue to accrete matter from the disc (e.g. \citep{rappaport+_04}).

\subsection{Model I }\label{sec:narrow}

The star is thought to be in the spin-down state for $\rA > \rco$ as the in-flowing disc matter is expelled from the system by the rotating magnetic field lines \citep{illarionov+_75}. \cite{ertan_17} set fort the condition that the magnetic torque acting on the in-flowing matter has to accelerate it to the speed of the field lines within the interaction time-scale in order to sustain a steady propeller. This condition leads that the star can accrete when $\rin \sim \rco$ even for $\rA \gg \rco$. In the case that disc-magnetosphere interaction occurs along a narrow boundary region, assuming the strength of the dipole field does not change significantly through the boundary, the total torque acting on the star by the disc can be written as 

\be 
\label{eq:torque_narrow}
\Gamma_{\mathrm{SD, 1}} = -\frac{\mu^{2}}{r_{\mathrm{in}}^{3}}\left(\frac{\Delta r}{r_{\mathrm{in}}}\right)
\ee

where $\Delta r$ denotes the boundary width. As predicted by \cite{ertan_17}, the SD torque strongly depends on the inner disc radius, 

\begin{eqnarray} \nonumber \label{eq:rin}
\lefteqn{R_{\mathrm{in}, \max }^{25 / 8}\left(1-R_{\mathrm{in}, \max }^{-3 / 2}\right) \simeq } \\
 & & 2 \times 10^{39}~\alpha_{-1}^{2 / 5}~M_{1.4}^{-7 / 6}~\dot{M}_{16}^{-7 / 20}~\mu_{30}~P^{-13 / 12} 
\end{eqnarray}

\citep{ertan_18}, where $\dot{M}_{16}$ is the mass accretion rate on to the star in the unit of $10^{16}$~g~s$^{-1}$, $\alpha_{-1}$ is the viscosity parameter for the disc in the unit of 0.1, $M_{1.4}$ is the mass of the neutron star in the unit of $1.4 \Msun$ and $\mu_{30}$ is the magnetic dipole moment of the neutron star in the unit of $10^{30}$~G~cm$^{3}$. The strength of the dipole field on the surface of the star is $B = \mu / R^3$. $R_{\mathrm{in}, \max }$ denotes the maximum inner disc radius (in the unit of $\rco$) for sustaining steady propeller. The inner disc radius should be close to but less than the maximum inner disc radius such that $R_{\mathrm{in}, \max } = (\rin/\eta)/\rco$ where $\eta$ is the parameter associated with the efficiency of the propeller. 

When $\rin < \rco$, we take $\rin = \rco$ in this model and the source is in the accretion phase where we assume all of the matter arriving at the inner disc flows on to the star. The X-ray luminosity is found by equation (\ref{eq:Lx}). In this phase, it is likely to observe the pulsed X-ray emission produced at the regions close to magnetic poles of the neutron star depending on the orientation of magnetic and rotation axes. We should note that the observed $\Lx$ and $\Pdot$ of \4U~ implies that the system has always been in the accretion phase even for $\rin$ estimated from equation (\ref{eq:rin}) while the total torque acting on the star can be negative and positive. Whether the system is in the SU or SD regimes can be determined by $\Mdotin, B, P$ for the given $M, \alpha$ and $\eta$.  

\subsection{Model II }\label{sec:wide}

\begin{figure}
\centering
\includegraphics[width=1.\columnwidth,angle=0]{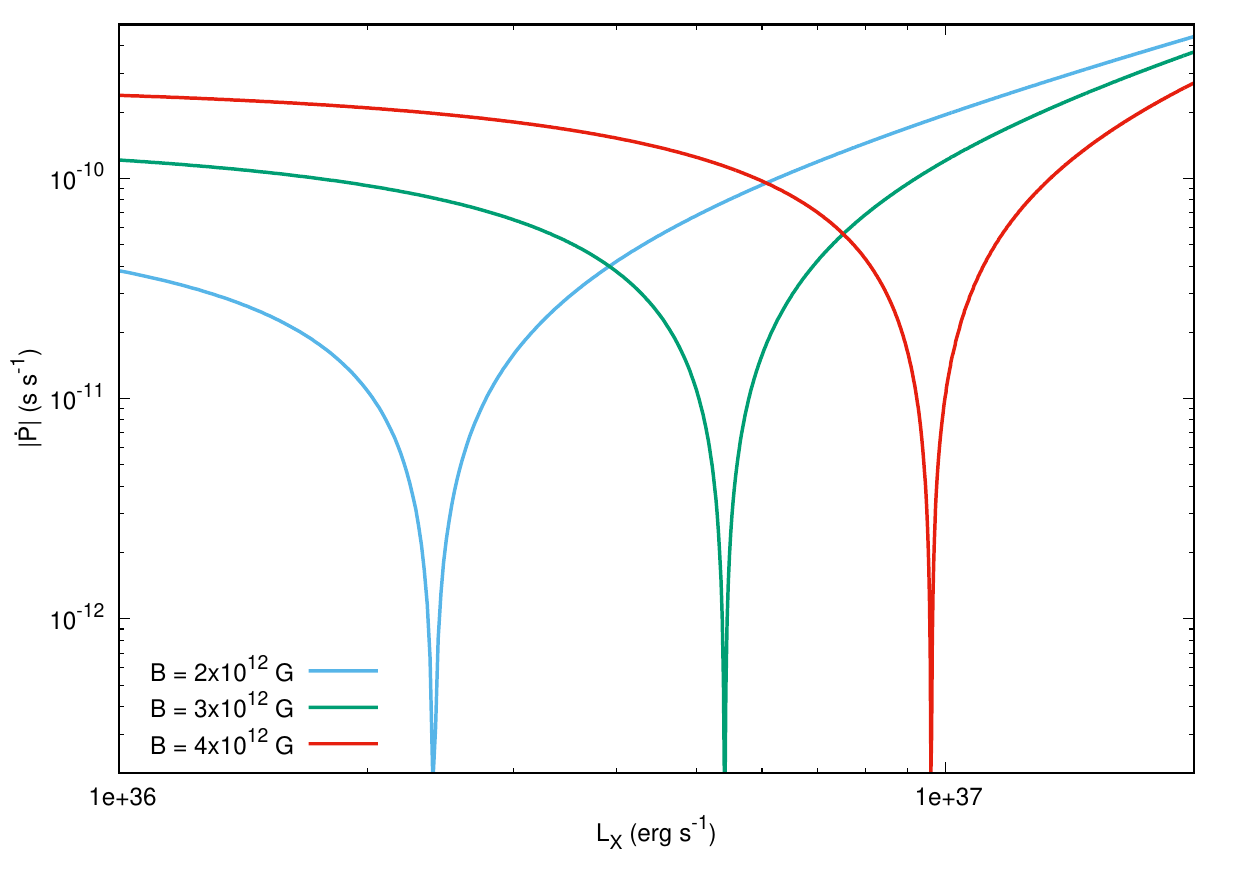}
\caption{Illustrative model curves indicating the change of the magnitude of $\Pdot$ by $\Mdot$ for three different magnetic field strengths given in the bottom left corner. The curves are produced by using equation (\ref{eq:torque_wide}) for the spin-down torque. The critical $\Lx$ for the torque reversal is determined by $B$. 
}

\label{fig:B}
\end{figure}

The spin-down disc torque acting on the neutron star can also be found by integrating the magnetic torque from $\rA$ to $\rco$ \citep{erkut_alpar_04}, 

\be
\Gamma_{\mathrm{SD, 2}} \simeq-\int_{r_{\mathrm{co}}}^{r_{\mathrm{A}}} r^{2} B_{\phi}^{+} B_{z} d r
\ee

where $B_{z} \simeq-\mu_{*} / r^{3}$ is the z-component of the dipolar magnetic field of the neutron star in the cylindrical coordinates. $B_{\phi}^{+}=\gamma_{\phi} B_{z}$ is the azimuthal component of the magnetic field on the surface of the disc and $\gamma_{\phi}$ denotes the azimuthal pitch. Due to the very sharp radial dependence of the magnetic pressure, most of the contribution to the spin-down torque comes from the radii close to $\rco$. In this case, taking $\gamma_{\phi} = 3/2$, the spin-down torque can be written as

\be 
\label{eq:torque_wide}
\Gamma_{\mathrm{SD, 2}} = I \dot{\Omega}_{*}=\frac{1}{2} \dot{M}_{\mathrm{in}}\left(G M r_{\mathrm{A}}\right)^{1 / 2}\left(1-\omega_{*}^{2}\right)
\ee

where $\dot{M}_{\mathrm{in}}$ is the rate of the mass inflow at $\rin$, $I$ is the moment of inertia of the star and $\omega_{*}$ is the fastness parameter defined by the ratio of $\rA$ to $\rco$, $\omega_{*}=\left(r_{\mathrm{A}} / r_{\mathrm{co}}\right)^{3 / 2}$. The critical fastness parameter, above which accretion is expected to be halted, is still controversial. We would like to draw attention to the point that in the accretion phase, equation (\ref{eq:torque_wide}) is found to be independent of $\dot{M}_{\mathrm{in}}$. In other words, spin-down torque is constant in the accretion phase. However, spin-up torque acting on the star, due to the angular momentum of the matter being accreted, (equation \ref{eq:su_torque}) causes varying total torque by $\dot{M}_{\mathrm{in}}$. There exist a particular regime that SU and SD torques are comparable. In this regime, the total torque decreases as $\dot{M}_{\mathrm{in}}$ increases but the star is still in the spin-down regime. Above a critical $\dot{M}_{\mathrm{in}}$, the total torque acting on the star switches its sign, i.e. the system makes a transition to the SU phase. In Fig. \ref{fig:B}, we show three model curves with different field strengths to indicate that the critical $\Lx$ where $\Pdot$ changes its sign strongly depends on the value of the magnetic field. The regimes where $\Pdot$ changes its sign while the star carries on accreting matter exist also for the Model I described in Section \ref{eq:torque_narrow}. So far, we have been dealing with the aligned rotator cases where the magnetic axis of the neutron star is parallel to its rotation axis. In the following section, we present a description of the model for a non-aligned configuration.   

\subsection{Model II + oblique rotator}\label{sec:oblique}
 
\cite{perna+_06} investigated the interaction between magnetosphere and disc for an oblique rotator (the angle between the spin axis of the pulsar and the disc-plane is $\chi \ne 0$). They supposed that the total torque is dominated by the material torque, however, they did not take into account the magnetic spin-down torque through the disc-magnetosphere boundary. Here, we additionally calculate the effect of spin-down torques but do not consider the re-joining of the ejected matter into the outer disc for the sake of simplicity in our model. \cite{perna+_06} predicted that, even if the mass supply from companion to the accretion disc is constant, cyclic transitions between SU and SD states can occur for larger inclination angles than a critical value. 


In cylindrical coordinates, the magnetic dipole field of an oblique rotator can be defined by 

\be
B^{2}=\frac{\mu^{2}}{r^{6}}\left[1+3(\sin \chi \sin \phi)^{2}\right]
\ee 

\citep{jetzer+_98, campana+_01} which leads magnetospheric radii of the elliptical form in the disc plane

\be
\rM (\phi) = \rA \left[1+3(\sin \chi \sin \phi)^{2}\right]
\ee

where $\phi$ is the azimuthal angle. In the co-rotating magnetosphere of the star, Kelvin-Helmholtz instability enables the material in the Keplerian disc to fill the empty regions within a short-timescales enough \citep{perna+_06}. The inner disc can, therefore, remains in steady contact with the magnetosphere. The speed of the field lines should be greater than the local escape velocity, i.e. $\rin > 1.26~\rco$, to enable the ejection of disc-material (at $\rin$) from the system by propeller mechanism. In the oblique rotator case, as $\rin$ is expected to change by the inclination angle $\chi$, the system could be in the partial propeller regime. It is possible that the partial azimuthal region is in the propeller phase while the remaining region can not expel the matter from the system. 

To take into account the effect of an inclined rotator, we employ a simple 2-D model involving the disc torques braking the star. Along the azimuthal angle grid, $\phi + \Delta \phi$, we integrate the amount of mass accreted and SU-material-torque and SD-disc-torque while assuming that all the incoming disc material is ejected for $\rM > \rco$. We implement the oblique case into Model II where disc-torques are integrated from $\rco$ to $\rA$ to compare with the our predictions for the aligned magnetic and rotation axis of the star.     

\begin{figure}
\centering
\includegraphics[width=1.\columnwidth,angle=0]{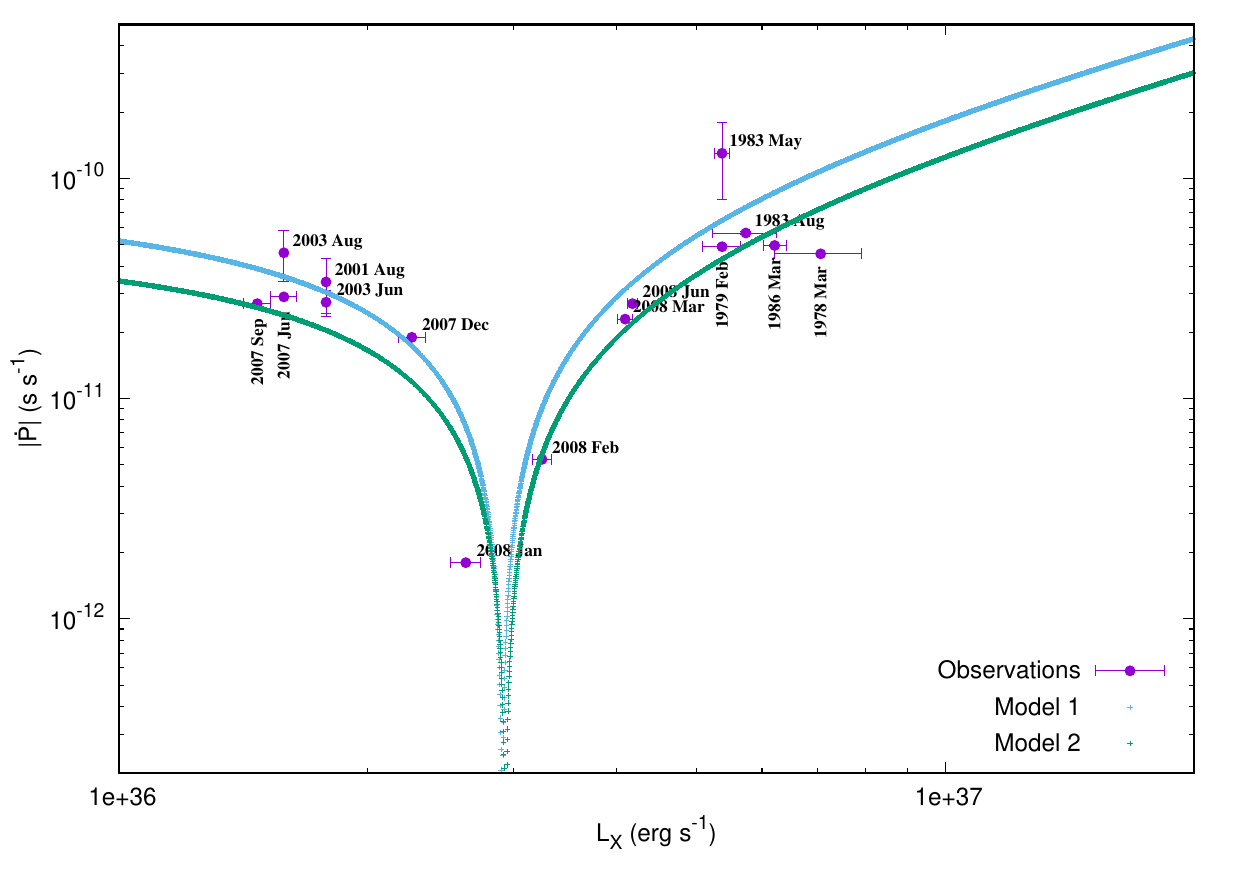}
\caption{Two model curves qualitatively fitting to the observed $\Lx$ and $\Pdot$ data of 4U 1626--67 for $d = 5$~kpc, $M = 1.4~\Msun$ and $R = 10$~km. The data points with error bars indicates the observed values where $\Lx$ is the bolometric luminosity \citep{takagi+_16}. The upper (blue) curve is calculated by Model II with $B =  2.2 \times 10^{12}$~G. The lower (green) curve is calculated by Model I with $\Delta r = 0.15$ and $B =  2.2 \times 10^{12}$~G.
}

\label{fig:5kpc}
\end{figure}

\section{Results }\label{sec:results}

\begin{figure}
\centering

\includegraphics[width=1.\columnwidth,angle=0]{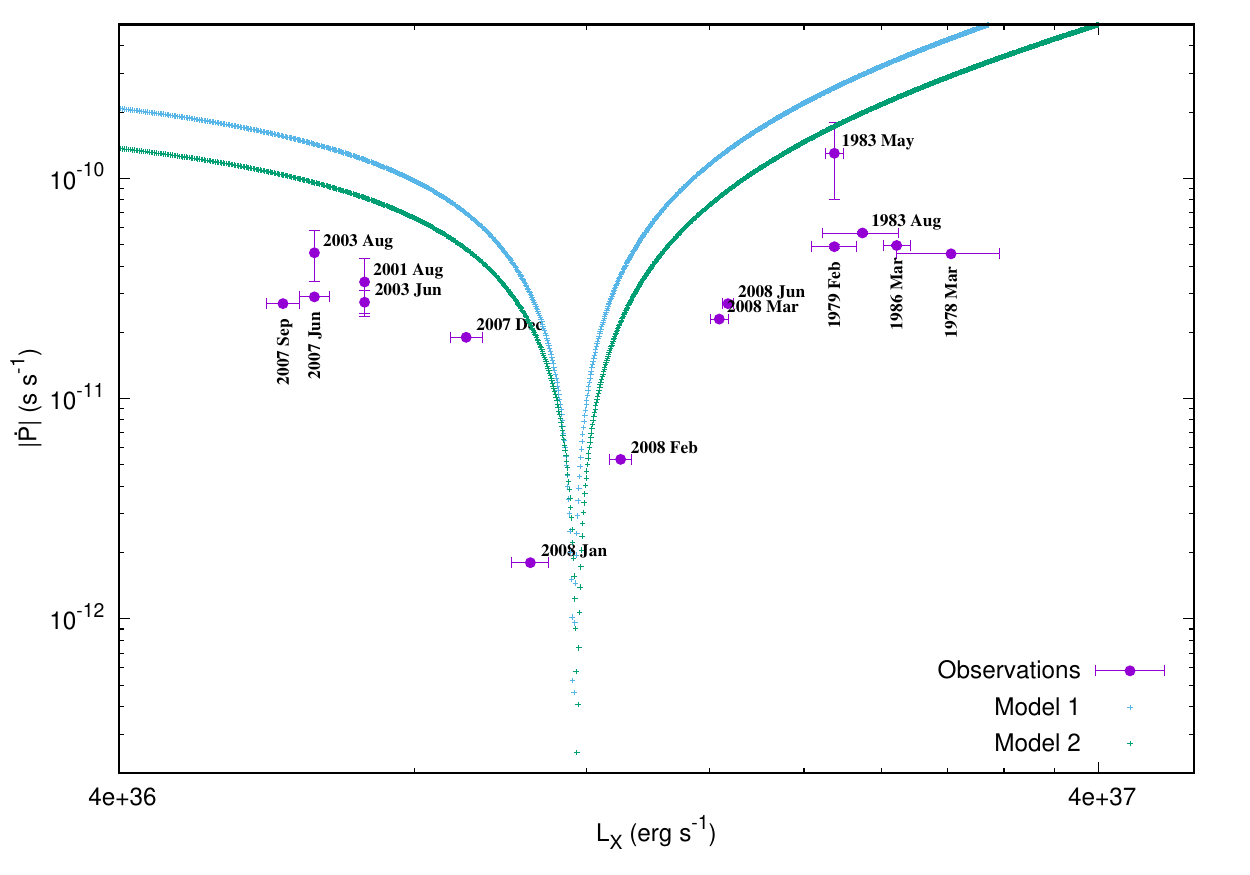}
\caption{Two illustrative model curves for comparing the observed $\Lx$ (for $d = 10$~kpc) and $\Pdot$ data of 4U 1626--67, with $M = 1.4~\Msun$ and $R = 10$~km. The predictions of the models are not consistent with the observed properties of the source for large distances.
}
\label{fig:10kpc}
\end{figure}

We investigated the torque and X-ray luminosity measurements of 4U 1626--67 performed at various time intervals for $\sim 40$~ years. We adopted the data presented by \cite{takagi+_16} and considered three different approaches in order to fit the data. The X-ray flux was converted into (0.5--100) keV energy band by \cite{takagi+_16} with the spectral model in \cite{camero-arranz+_12}. The $\Lx$ and $\Pdot$ observations in both spin-up and spin-down phases of 4U 1626--67 were examined by the aligned and the orthogonal rotator assumptions. Figs. \ref{fig:5kpc} and \ref{fig:10kpc} show our results for the aligned rotator, for the distances $d = 5$~kpc and $d = 10$~kpc. In all our calculations, we adopted the conventional mass and radius of neutron star, $M = 1.4~\Msun$ and $R = 10$~km. Within the possible range of the source distance, 5--13~kpc \citep{chakrabarty_98, takagi+_16}, we tried different distances to fit the data and found that the lowest distance, $d = 5$~kpc, produces the best model curves. We also present our results for $d = 10$~kpc in order to indicate that fitting quality decreases for larger distances.  

The upper and lower model curves in Figs. \ref{fig:5kpc} and \ref{fig:10kpc} are produced by employing Model I and Model II, respectively. The data points with error bars indicate $\Lx$ and $\Pdot$ measurements for the years presented next to the data points on the figures. Y-axes indicate the magnitude of $\Pdot$ given in the log-scale. The dips appearing in the $\Pdot-\Lx$ curves remark the transition luminosity between SU and SD phases. The left-hand side of the dip corresponds to the SD phase with a positive $\Pdot$. For greater luminosities on the right-hand side of the dip, the source is in the SU phase and the $\Pdot$ reversed its sign at the most bottom of the dip which is not seen on the figures. We would like to point out that the two data points, January 2008 and February 2008, were measured when the source was in the torque reversal epoch. These two data points determine the transition location and constrain the width of the dip.

In Fig \ref{fig:5kpc}, the observed $\Lx$ and $\Pdot$ of 4U 1626--67 are produced with the dipole field strengths $B =  2.2 \times 10^{12}$~G and $B =  3.2 \times 10^{12}$~G by using equations \ref{eq:torque_wide} and \ref{eq:torque_narrow}. The boundary width of magnetosphere-disc interaction $\Delta r = 0.15$ is used for equation \ref{eq:torque_narrow}. The less efficient torque definition given by equation (\ref{eq:torque_narrow}) could account for the two data point during the transition in 2008 and the data in the spin-up phase better. However, the torque formula in equation (\ref{eq:torque_wide}) could fit the data in the spin-down phase better than that is in equation (\ref{eq:torque_narrow}).

\begin{figure}
\centering
\includegraphics[width=1.\columnwidth,angle=0]{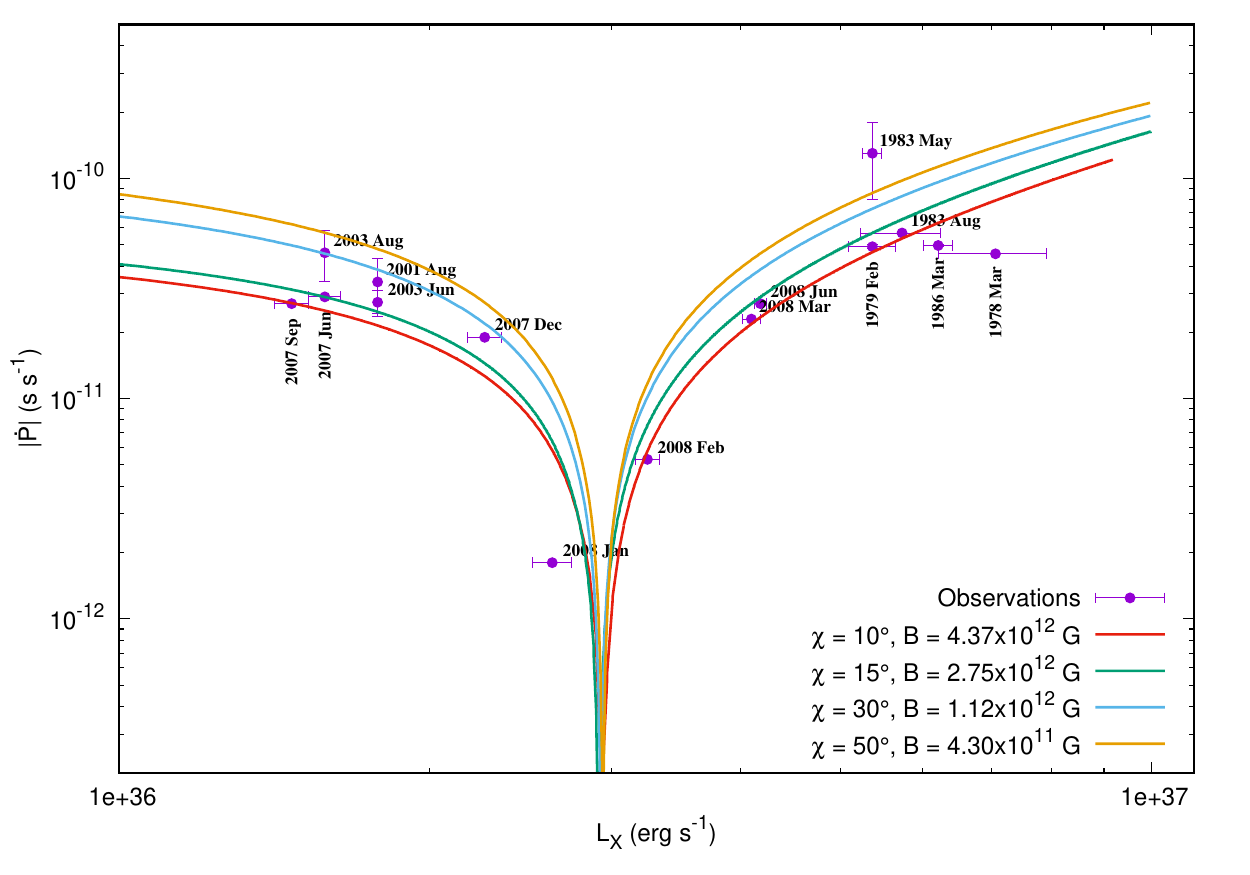}
\caption{The model curves for different angles between the rotation axis and the magnetic axis, $\chi$, with the field strengths given in the bottom right of the figure. There is a unique $B$ value for each $\chi$ which could result in the torque reversing at the same critical $\Lx$. 
}
\label{fig:obliqe}
\end{figure}

Employing the SD-torque expression that is found by integrating disc torques from $\rco$ to $\rA$ and assuming an oblique rotator for the star, we found the similar results $\Pdot-\Lx$ curves for the properties of \4U. However, depending on the angle between the magnetic axis and rotation axis, we estimated different magnetic field strengths of the star. We present the model results in Fig. \ref{fig:obliqe} and Table \ref{table:1}. We found that the best result consistent with the observed properties of \4U could be found with $\chi \sim 15$ and $B = (2-3) \times 10^{12}$~G for $d = 5$~kpc. For this parameters, the system is found to be in such an interesting regime that a small increase in $\Mdot$ leads a substantial change in $\Lx$, e.g. $\Lx = 2 \times 10^{36}$~erg~s$^{-1}$ is produced with $\Mdot \simeq 8 \times 10^{16}$~g~s$^{-1}$ while $\Mdot \simeq 9.3 \times 10^{16}$~g~s$^{-1}$ gives $\Lx = 6 \times 10^{36}$~erg~s$^{-1}$. Therefore, the observed range of $\Lx$ from \4U does not require a significant change of mass accretion rate from the companion in this model.


\begin{table}
\label{table:1}
\caption{The required amount of change in $\Mdot$ for a 3-fold increase in $\Lx$ for the model curves presented in Fig. \ref{fig:obliqe}. Each line represents the different angles between the rotation axis and the magnetic axis. $\Mdot_1$ and $\Mdot_2$ are the mass accretion rates which give $\Lx = 2 \times 10^{36}$~erg~s$^{-1}$ and $\Lx = 6 \times 10^{36}$~erg~s$^{-1}$. The magnetic field strengths that could produce each model curve is listed in the last column.  }
\centering
\begin{tabular}{c|ccc}
\hline 
				& $\Mdot_1$ (g~s$^{-1}$) & $\Mdot_2$ (g~s$^{-1}$) & $B$ (G) \\
\hline 
$\chi = 10^{\circ}$		& $1.97 \times 10^{17}$ & $2.00 \times 10^{17}$	  & $4.37 \times 10^{12}$	\\
$\chi = 15^{\circ}$		& $8.02 \times 10^{16}$ & $9.34 \times 10^{16}$	  & $2.75 \times 10^{12}$	\\
$\chi = 30^{\circ}$		& $2.79 \times 10^{16}$ & $5.32 \times 10^{16}$	  & $1.12 \times 10^{12}$	\\
$\chi = 50^{\circ}$		& $2.00 \times 10^{16}$ & $4.40 \times 10^{16}$	  & $4.30 \times 10^{11}$	\\
\hline 

\end{tabular}
\end{table}

   
\section{Discussion and Conclusion} \label{sec:dis_conc}

The cyclotron line feature of 4U 1626--67 at $\sim 37$~keV indicates that the strength of the magnetic dipole field is $B \simeq 3 \times 10^{12}$~G at the surface of the star \citep{orlandini+_98, dai+_17}. The torque reversal did not accompany a change in cyclotron line energy while X-ray flux increased a factor of $\sim 3$ \citep{camero-arranz+_12}. Also employing the beat frequency model, \cite{orlandini+_98} estimated that the magnetic field is $B \sim (2-6) \times 10^{12}$~G, comparable to the field strength indicated by the cyclotron line feature. The model curves that we showed to be good candidates to account for the observed $\Pdot$ and $\Lx$ of the source, are produced with $B \simeq 2-3 \times 10^{12}$~G that is in line with those predictions. 

During the spin-down phase of 4U 1626--67, \cite{chakrabarty_98} detected 48 mHz quasi-periodic oscillation (QPO) in both X-ray and optical bands. \cite{raman+_16} analysed the 2014-data of the source and detected a weak $\sim 48$~mHz optical QPO but could not detect any QPO in the corresponding X-ray data. The QPO frequency is thought to be associated with the angular velocity of the pulsar and Keplerian angular velocity at the inner disc as such in the beat frequency model, $\omega=\Omega_{\mathrm{K}}\left(r_{\mathrm{in}}\right)-\Omega_{*}$ \citep{alpar+_85}. Therefore, investigating the evolution of the QPO frequency of accreting sources with varying mass accretion rates is of great importance to trace the actual mass accretion rate. Correct determination of the accretion rate is essential in order to obtain a proper torque model accounting for the strange $\Lx$ and torque behaviors.

\cite{bildsten+_97} raised the question \textit{How does the companion star know just how to change its mass transfer rate so that transitions between two torques of the same sign never occur?} There is no convincing answer to this question yet. According to our results, this may be caused by the selection effect. The sharp change of $\Pdot$, when $\Mdot$ is close to the critical mass accretion rate, could make it harder to observe $\Pdot$ within this range. However, as the magnitude of $\Pddot$ decreases sufficiently when $\Mdot$ is far from the SU/SD transition point, measuring $\Pdot$ in both SU and SD phases is more likely for the source far from the transition. This could be one possibility why $|\Pdot|$ is in the same order in both phases.  

The models that are built upon the disc torque via interaction by the dipole field lines, require fine-tuned alternation of the mass accretion rates to reproduce the observed torques of similar magnitude but with opposite signs. \cite{makishima+_88} proposed the idea that the SU-SD transition of GX 1+4 might be caused by the disappearance of the prograde disc and subsequent formation of the wind-fed retrograde disc. The alternations in the rotation direction of the discs could lead the torques with different signs but similar magnitudes. The models developed upon the retrograde disc suggestion could explain the inverse proportionality of the X-ray luminosity and spin-up torque as well as the positive correlation of spin-down torque and the X-ray luminosity of GX 1+4 \citep{chakrabarty+_97, nelson+_97}. It is, however, unlikely that Roche lobe overflow accretion can result in a retrograde disc. Unlike GX 1+4, 4U 1626--67 has a very low-mass companion. Therefore, imposing retrograde accretion disc is inconvenient to explain the observed torque inversion of 4U 1626--67.

As an alternative model, warped and precessing inner disc driven by the radiation pressure in the high X-ray luminosity systems or by the tidal forces caused by the companion, could be responsible for the curious torque behaviors \citep{vanKerkwijk+_98, wijers_pringle_99}. In this model, torque reversals of 4U 1626--67 was anticipated to occur when the inner disc tilted more than $90^{\circ}$ to the unperturbed disc plane. The outer regions of the disc do not necessarily be tilted more than $90^{\circ}$. In other words, the disc is partially counter-rotating while its outer part remains prograde. Investigation of the precessing disc and its stability check is beyond our scope in this work. We think that more detailed numerical simulations are requisite in order to favour those kinds of models in addition to richer data from the sources experiencing torque reversals.

\section*{Acknowledgements}
We are grateful to {\"U}. Ertan for stimulating discussions. We thank D. Altamirano and J. P\'{e}tri for useful comments. We acknowledge research support from T\"{U}B{\.I}TAK (The Scientific and Technological Research Council of Turkey) through the postdoctoral research programme (BİDEB 2219) and CEFIPRA grant IFC/F5904-B/2018.


\bibliographystyle{mn2e}
\bibliography{benli.bib}

\label{lastpage}

\end{document}